\documentclass[11pt]{article}
\usepackage{amsmath}
\usepackage{amssymb}
\usepackage{amsfonts}
\usepackage{eucal}
\usepackage{epsfig}
\newcommand{\be}{\begin{equation}}
\newcommand{\bea}{\begin{eqnarray}}
\newcommand{\eea}{\end{eqnarray}}
\newcommand{\ba}{\begin{array}}
\newcommand{\ea}{\end{array}}
\newcommand{\ee}{\end{equation}}

\expandafter\ifx\csname mathbbm\endcsname\relax

\else

\fi

\renewcommand{\baselinestretch}{1}

\begin{document}

\title{An analytical formulation for $\phi^{4}$ field-potential dynamics}  \author{Arash Ghahraman \\ ar$_{-}$gh51@stu-mail.um.ac.ir \\Kurosh Javidan
\\javidan@um.ac.ir \\Department of physics, Ferdowsi university of Mashhad\\ 91775-1436 Mashhad Iran }
\setlength{\parindent}{0mm} 
\renewcommand{\baselinestretch}{1} 
\newcommand{\ph}{\vec{\phi}} \newcommand{\pha}{\phi_{a}}
\newcommand{\dmu}{\partial_{\mu}} \newcommand{\umu}{\partial^{\mu}}
\newcommand{\dnu}{\partial_{\nu}} \newcommand{\unu}{\partial^{\nu}}
\newcommand{\di}{\partial_{i}} \newcommand{\ui}{\partial^{i}}
\renewcommand{\dj}{\partial_{j}} \newcommand{\uj}{\partial^{j}} \hoffset =
-2cm \textwidth = 170mm
\maketitle
\abstract

An analytical model for adding a space dependent potential to the $\phi^{4}$ field is presented, by constructing a collective coordinate for solitary solution of this model. Interaction of the $\phi^{4}$ field with a delta function potential barrier and also delta function potential well is investigated. Most of the characters of the interaction are derived analytically while they are calculated by other models numerically. We will find that the behaviour of the solitary solution is like a point particle 'living' under the influence of a complicated potential which is a function of the field initial conditions and the potential parameters. 

\section{Introduction}

Dynamical evolution of a field in the presence of an external potential, in which case the parameters of the model are functions of space is an important phenomenon from the mathematical point of view and also because of its applications. An external potential can be added to the equation of motion as perturbative terms [1, 2]. These effects also can be taken into account by making some parameters of the equation of motion to be function of space or time [3, 4].  The external potential also can be added to the field through the metric of background space-time [5, 6]. This method is mainly suitable for nonlinear field theories contain solitonic solutions.

As is well known, when waves scatter on a potential, they can be partly reflected and partly transmitted. For the fields with solitonic solutions, the situation is more complicated as solitons cannot split and thus must either bounce, pass through or become trapped inside the potential. This behaviour is very sensitive to the value of all the parameters of the model as well as to the initial conditions for the scattering. Most of the researches are in base of numerical studies because such these systems are generally non-integrable. So it is clear that we need suitable models with analytic solutions to test the validity of such phenomenon and predict their behaviour.

In this paper an analytical model for the interaction of $\phi^{4}$ field with an external potential is presented. This method can be used for other field theories and from this viewpointin, we will present an example for explaining the method. However the results are perfectly valid for the $\phi^4$ field theory. So a model for the $\phi^{4}$ field in an space dependent potential is presented in section 2. The analytic model is introduced and will be solved in section 3. The results for the field-potential barrier system are presented in section 4. In section 5 field-potential well system is discussed. Some conclusion and remarks will be presented in section 6.

\section{$\phi^{4}$ field and a space dependent potential }
The general form of the action in an arbitrary metric is:
\begin{equation}\label{I}
I=\int{{\cal L}(\phi , \partial_{\mu}\phi)\sqrt{-g}d^{n}x dt }
\end{equation}
where "g" is the determinant of the metric $g^{\mu \nu} (x)$. Energy density of the "field + potential" can be found by varying "both" the field and the metric \cite {r7}. For the lagrangian of the form
\begin{equation}\label{l}
{\cal L}=\frac{1}{2}\partial_{\mu}\phi\partial^{\mu}\phi-U(\phi)
\end{equation}
the equation of motion becomes \cite {r7,r8}
\begin{equation}\label{Em}
\frac {1}{\sqrt{-g}}\left (\sqrt{-g}\partial_{\mu}\phi\partial^{\mu}\phi+\partial_{\mu}\phi\partial^{\mu}\sqrt{-g}\right )+\frac {\partial U(\phi)}{\partial \phi}=0
\end{equation}
One can add a space dependent potential to the lagrangian of the system by introducing a suitable nontrivial metric for the back ground space-time, without missing the topological boundary conditions \cite {r6,r7}. In other words, the metric carries the introduced potential.
The suitable metric in the presence of a weak potential $v(x)$ is \cite {r5,r6,r7}:
\begin{equation}\label{metric}
g_{\mu \nu}(x)\cong\left(
\begin{array}{clrr} 1+V(x) & 0 \\ 0 & -1
\end{array}\right)
\end{equation}
The equation of motion (3) (describes by Lagrangian (2)) in the background space-time (4) is
\begin{equation}\label{Eme}
\left ( 1+V(x)\right )\frac {\partial^{2}\phi}{\partial t^2}-\frac {\partial^{2}\phi}{\partial x^2}-\frac {1}{2\left|1+V(x)\right|}\frac {\partial V(x)}{\partial x}\frac {\partial \phi}{\partial x}+\frac {\partial U(\phi)}{\partial \phi}=0
\end{equation}
For the $\phi^{4}$ model, we have $U(\phi)=\left(\phi^{2}-1\right)^{2}$. The $\phi^{4}$ model has solitary wave solution 
\begin{equation}\label{sol}
\phi (x,t)= tanh \left(\sqrt{2} \frac{ x-X(t)}{\sqrt{1-v^{2}}}\right)
\end{equation}
where $X(t)=x_{0}-vt$ and $v$ is the velocity of the solitary wave. 
By inserting the solution (6) in the lagrangian (2) and using the metric (4), with adiabatic approximation \cite {r1,r2} we have
\begin {equation} \label {le1}
{\cal L}=\sqrt{1+v(x)}\left(\left(1+v(x)\right)\dot{X}^{2}-2\right)sech^{4}\left(\sqrt{2}\left(x-X(t)\right)\right) 
\end {equation}
For the weak potential $v(x)$ (7) becomes
\begin {equation} \label {lea}
{\cal L}\approx\left(\left(1+\frac{3}{2}v(x)\right)\dot{X}^{2}-2\left(1+\frac{1}{2}v(x)\right)\right)sech^{4}\left(\sqrt{2}\left(x-X(t)\right)\right) 
\end {equation}

\section{Collective coordinate variable }
The center of a soliton can be considered as a particle, if we look at this variable as a collective coordinate. The collective coordinate could be related to the potential by using the above model. This model is able to give us an analytic solution for most of the features of the soliton-potential system. For the lagrangian (8) X(t) remains as a collective coordinate if we integrate (8) over variable x
\begin {equation} \label {L}
L=\int{{\cal L}dx}=\frac{2\sqrt{2}}{3}\dot{X}^{2}-\frac{4\sqrt{2}}{3}+\left(\frac{3}{2}\dot{X}^{2}-1\right)\int{sech^{4}\left(\sqrt{2}\left(x-X(t)\right)\right)v(x)dx}
\end {equation}
If we take the potential $v(x)=\epsilon \delta(x)$ then (9) becomes
\begin {equation} \label {lap}
L=\frac{2\sqrt{2}}{3}\dot{X}^{2}-\frac{4\sqrt{2}}{3}+\left(\frac{3}{2}\dot{X}^{2}-1\right)\epsilon sech^{4}\left(\sqrt{2}X(t)\right)
\end {equation}
The equation of motion for the variable X(t) is derived from (10)
\begin {equation} \label {EmX}
\left(\frac{2\sqrt{2}}{3}+\frac{3\epsilon}{2} sech^{4}\left(\sqrt{2}X\right) \right)\ddot{X}-\sqrt{2}\epsilon tanh(\sqrt{2}X)sech^{4}\left(\sqrt{2}X\right)\left(3\dot{X}^{2}+2\right)=0
\end {equation}
 
The above equation shows that the peak of the soliton moves under the influence of a complicated force which is a function of its position and its velocity. If $\epsilon>0$ we have a barrier and $\epsilon<0$ creates a potential well. Fortunately equation (11) has an exact solution as follows
\begin {equation} \label {Xdot1}
\frac{3\dot{X}^2 +2}{3\dot{X}_{0}^2 +2}=\frac{\frac{2\sqrt{2}}{3}+\frac{3\epsilon}{2}sech^{4}\left(\sqrt{2}\left(X_{0}\right)\right)}{\frac{2\sqrt{2}}{3}+\frac{3\epsilon}{2}sech^{4}\left(\sqrt{2}\left(X\right)\right)}
\end {equation}
where $X_{0}$ and $\dot{X}_{0}$ are initial position and initial velocity respectively. The energy of the soliton in the presence of the potential $v(x)=\epsilon \delta(x)$ becomes
\begin {equation} \label {Energy}
E=\left(\frac{2\sqrt{2}}{3}+\frac{3\epsilon}{2}sech^{4}\left(\sqrt{2}X\right) \right)\dot{X}^2+\frac{4\sqrt{2}}{3}+\epsilon sech^{4}\left(\sqrt{2}X\right)
\end {equation}
Subtituting $\dot{X}$ in (13) from (12) shows that the energy is a function of initial conditions $X_{0}$ and $\dot{X}_{0}$ only. Therefore the energy of the sytem is conserved. Some features of the soliton-potential dynamics can be investigated using equations (12) and (13) analytically which are discussed in the following.

\section{Potential barrier}
Suppose that a potential barrier of the height $\epsilon$ is located at the origin. When the soliton is far from the center of the potential ($X\rightarrow\infty$) (13) reduces to $E=\frac{2\sqrt{2}}{3}\dot{X}_{0}^2+\frac{4\sqrt{2}}{3}$. It is the energy of a particle with a mass of $\frac{4\sqrt{2}}{3}$ and a velocity of $\dot{X}_{0}$. A soliton with a low velocity reflects back from the barrier and a high energy soliton climbs over the barrier and passes over it. So we have a critical value for the velocity of the soliton which separates these two situations. The energy of a soliton in the origin (X=0) comes from (13)$E(X=0)=\left( \frac{2\sqrt{2}}{3}+\frac{3\epsilon}{2}\right)\dot{X}_{0}^2+\frac{4\sqrt{2}}{3}+\epsilon$. The minimum energy for a soliton in this position is $E=\frac{4\sqrt{2}}{3}+\epsilon$. On the other hand, a soliton which comes from the infinity with initial velocity $v_{c}$ has the energy of $E\left(X=\infty\right)=\frac{2\sqrt{2}}{3}v_{c}^2+\frac{4\sqrt{2}}{3}$. Therefore it can pass through the barrier if $v_{c}>\sqrt{\frac{3\sqrt{2}}{4}\epsilon}$. The same result is derived by substituting $\dot{X}=0$, $\dot{X}_{0}=v_{c}$, $X_{0}=\infty$ and $X=0$ in (12).

If the soliton is located at some position like $X_{0}$ (which is not necessary infinity) the critical velocity will not be $\sqrt{\frac{3\sqrt{2}}{4}\epsilon}$. The soliton can pass over the barrier if the soliton energy is greater than the energy of a static soliton at the top of the barrier. So a soliton in the initial position $X_{0}$ with initial velocity of $\dot{X}_{0}$ has the critical initial velocity if its velocity becomes zero at the top of the barrier $X=0$. Consider a soliton with initial conditions of $X_{0}$ and $\dot{X}_{0}$. If we set $X=0$ and $\dot{X}=0$ in equation (12) then $v_{c}=\dot{X}_{0}$. Therefore we have 
\begin {equation} \label {vc}
v_{c}=\sqrt{\frac{\epsilon\left(1-sech^{4}\left(\sqrt{2}X_{0}\right)\right)}{\frac{2\sqrt{2}}{3}+\frac{3\epsilon}{2}sech^{4}\left(\sqrt{2}x_{0}\right)}}
\end {equation}

When the initial velocity is less than the $v_{c}$ then there exists a return point in which the velocity of the soliton is zero. For this situation we have
\begin {equation} 
\left(\frac{3}{2}\dot{X}_{0}^{2}+1\right)\left(\frac{2\sqrt{2}}{3}+\frac{3\epsilon}{2}sech^{4}\left( \sqrt{2}X_{0} \right) \right)-\frac{2\sqrt{2}}{3}=\frac{3\epsilon}{2}sech^{4}\left(\sqrt{2}X_{stop}\right)
\end {equation}
Therefore this model predicts that $sech^{4} \left(\sqrt{2}X_{stop}\right)$ is proportional to $sech^{4} \left(\sqrt{2}X_{0}\right)$. Also (15) shows that $sech^{4} \left(\sqrt{2}X_{stop}\right)$ is proportional to $\frac{1}{\epsilon}$. 

A soliton with initial conditions $X_{0}$ and $\dot{X}_{0}$ will go to the infinity after the interaction with a potential barrier. The final velocity of the soliton after the interaction is
\begin {equation} 
\dot{X}=\sqrt{\dot{X}_{0}^{2} +\frac{3\epsilon}{2\sqrt{2}}\left(\frac{3\dot{X}_{0}^2}{2}+1 \right)sech^4 \left(\sqrt{2}X_{0}\right)     }
\end {equation}
which is greater than the initial velocity $\dot{X}_{0}$.

Equations (12) and (13) show that the soliton finds its initial velocity after the interaction when it reaches its initial position. This means that the interaction is completely elastic. Numerical simulation are in agreement with above analytical results \cite {r7,r9,r10}.

\section {Soliton-well system}

The soliton-well system is very interesting problem. Suppose a particle moves toward a frictionless potential well. It falls in the well with increasing velocity and reaches the bottom of the well with its maximum speed. After that, it will climb the well with decreasing velocity and finally pass through the well. Its final velocity after the interaction is equal to its initial speed. 

Changing $\epsilon$ to $-\epsilon$ in (12) changes potential barrier to potential well. The solution for the system is
\begin {equation} \label {Xdot}
\frac{3\dot{X}^2 +2}{3\dot{X}_{0}^2 +2}=\frac{\frac{2\sqrt{2}}{3}-\frac{3\epsilon}{2}sech^{4}\left(\sqrt{2}\left(X_{0}\right)\right)}{\frac{2\sqrt{2}}{3}-\frac{3\epsilon}{2}sech^{4}\left(\sqrt{2}\left(X\right)\right)}
\end {equation}

There is not a critical velocity for a soliton-well system, but an escape velocity can be defined. A soliton with initial position $X_{0}$ reaches the infinity with a zero final velocity if its initial velocity is
\begin {equation} \label {Xdot0}
\dot{X}_{escape}=\sqrt{\frac{\epsilon sech^{4}\left(\sqrt{2}X_{0}\right)}{\frac{2\sqrt{2}}{3}-\frac{3\epsilon}{2}sech^{4}\left(\sqrt{2}X_{0}\right)}}  
\end {equation}
In other words, a soliton which is located in the initial position $X_{0}$ can escape to infinity if its initial velocity $\dot{X}_{0}$ is greater than the escape velocity $\dot{X}_{escape}$.
  
Consider a potential well with the depth of $\epsilon$ and a soliton at the initial position $X_{0}$ which moves toward the well with the initial velocity $\dot{X_{0}}$ smaller than the $\dot{X}_{escape}$. The soliton interacts with the potential and reaches a maximum distance $X_{max}$ from the center potential with a zero velocity and then come back toward the well. The soliton oscillates around the well with the amplitude $X_{max}$.  The required initial velocity to reach $X_{max}$ is found from (17) as
\begin {equation} 
\dot{X}_{0}=\sqrt{\frac{\epsilon \left(sech^{4}\left(\sqrt{2}X_{max}\right)-sech^{4}\left(\sqrt{2}X_{0}\right)\right)}{\frac{2\sqrt{2}}{3}-\frac{3\epsilon}{2}sech^{4}\left(\sqrt{2}X_{max}\right)}}  
\end {equation}

If the initial velocity is lower than the escape velocity the soliton oscillates around the well. The period of oscillation can be calculated numerically using equation (17).

\section{Conclusion and Remarks}

A model for the $\phi^{4}$ field-potential interaction has been presented. The model predicts a critical velocity for the soliton-barrier interaction as a function of initial conditions of the field and the potential characters. The model predicts specific relations between some functions of initial conditions and other functions of final state of the field during the interaction. An escape velocity has been derived for the field-well system. This model is able to explain most of the features of the system analytically. 

This model can be used for prediction the results of other field theories beside the $\phi^{4}$ model.


\end{document}